\documentclass[11pt,dvips]{article}
\usepackage{epsfig}
\usepackage{graphicx}
\usepackage{epic}
\usepackage{eepic}
\usepackage{amssymb}
\title{The Ising Model on a Dynamically Triangulated Disk with a Boundary Magnetic Field}
\author{Scott McGuire, Simon Catterall, Mark Bowick, Simeon Warner}
\date{\today}
\begin{document}

\begin{flushright}
SU-4240-720\\
LAUR-01-2140\\
\today
\end{flushright}
\vspace{0.5in}

\begin{center}
{\Large\bf The Ising Model on a Dynamically Triangulated Disk with a Boundary Magnetic Field}
\vspace{0.5in}

{\small Scott McGuire$^1$, Simon Catterall$^{1,3}$, Mark Bowick$^1$, Simeon Warner$^2$\\
	$^1$ Physics Department, 
        Syracuse University, 
        Syracuse, NY 13244 \\
        $^2$
	T-8, LANL,
	Los Alamos, NM 87545\\
        $^3$
        Corresponding Author
}

\end{center}


\begin{abstract}
    We use Monte Carlo simulations to study a dynamically triangulated
disk with Ising spins on the vertices and a boundary magnetic field.
For the case of zero magnetic field we show that the model possesses
three phases. For one of these the boundary length grows linearly with
disk area, while the other two phases are characterized by a boundary
whose size is on the order of the cut-off.  A line of continuous
magnetic transitions separates the two small boundary phases.  We
determine the critical exponents of the continuous magnetic phase
transition and relate them to predictions from continuum 2-d quantum
gravity.  This line of continuous transitions appears to terminate on
a line of discontinuous phase transitions dividing the small boundary
phases from the large boundary phase.  We examine the scaling of bulk
magnetization and boundary magnetization as a function of boundary
magnetic field in the vicinity of this tricritical point.
\end{abstract}

\section{Introduction}
    The dynamical triangulations approach to (Euclidean) 2-d
quantum gravity~\cite{KAZAKOV:ESM,DAVID:SQG,AJW:DTA} arose from
attempts to apply the Feynman path integral prescription for
quantizing field theories to gravity.  A straightforward
application of the prescription would involve integrating over all
2-d manifolds of a given topology. For generic matter
coupled systems this integration  cannot be done. The dynamical
triangulation method replaces the integral with a sum over
discretized manifolds composed of identical equilateral triangles
connected along their edges. In addition to including several
exactly solvable models, the approach is also amenable to computer
simulation using techniques from statistical physics. This is the
approach we have taken in this paper.

    Most work based on 2-d dynamical triangulations so far has been
restricted to closed topologies where Liouville theory provides many
predictions.  The use of a topology with a boundary allows us to probe
less well understood areas --- in this case the Ising model coupled to
a boundary magnetic field ($H$) and 2-d quantum gravity.  This has
been studied analytically~\cite{COT:BFA,COT:TIM,COT:AGA}.  The
boundary magnetic field can be seen as interpolating between the only
two conformally invariant boundary conditions: fixed ($H = \pm
\infty$) and free ($H = 0$).  Away from these special boundary
conditions, the model possesses the interesting property of being
conformally invariant in the bulk, but not at the boundary.  We
describe our model in more detail in Section~\ref{sec-model}.

    In principle, the model has four couplings. In the gravitational
sector, there are couplings to the disk area (number of triangles,$A$)
and to the boundary length ($l$).  In the matter sector, there are the
Ising coupling ($J$) and the boundary magnetic field.  We describe our
methods for simulating the model in Section~\ref{sec-methods}.  They
were initially tested against a strong coupling expansion of the
partition function.  This along with measurements of the disk
amplitude for zero Ising coupling are discussed in
Section~\ref{sec-tests}.  Both agree with predictions.

     For fixed disk area and zero boundary magnetic field, we have
mapped the resultant 2-d phase diagram.  We show evidence for three
phases and associated critical lines, and use finite size scaling to
extract estimates for critical exponents. These lines of transitions
appear to meet at a unique tricritical point at which the system
exhibits long range spin correlations and a non-trivial scaling of
boundary length with disk area.  This is presented in
Section~\ref{sec-phases}.

    As explained in Section~\ref{sec-boundary}, we have tracked this
tricritical point as a function of boundary magnetic field.  We show our
results concerning the scaling behavior of the model with non-zero
boundary magnetic field and test some of the conjectures contained
in~\cite{COT:BFA}. Finally, we describe possible future work.

\section{The Model} \label{sec-model}
    Our model consists of a 2-d triangulation composed of
equilateral triangles connected along their edges with
Ising~\cite{ISING} spins on the vertices.  We allow only combinatorial
triangulations; that is we exclude degenerate triangulations --- no
two triangles have the same three vertices, no triangle includes the
same vertex twice and each triangle may have only one neighbor per
edge.  We characterize a configuration by its area, $A$,
boundary length, $l$, and Ising spins, $\sigma_i$.  The action for
such a configuration is
\begin{equation}
    S = -J \sum_{\left<ij\right>} \sigma_i \sigma_j -H \sum_{k \in \partial} \sigma_k
        + \mu A + \nu l
\end{equation}
where $\left<ij\right>$ are nearest neighbor vertices, $\partial$
denotes the boundary of the disk, $J$ is the Ising coupling, and $\mu$
and $\nu$ are the cosmological constant and boundary cosmological
constant respectively.  The action contains no explicit gravitational
component because, in two dimensions, the Gauss-Bonnet theorem implies
that the Einstein-Hilbert action is a topological
invariant~\cite{DAVID:SQG}.  The grand canonical partition function is
\begin{equation}
    Z = \sum_{A,l} \sum_{t_{Al}} \sum_{\{\sigma_i\}} e^{-S}.
\end{equation}
where $\sum_{t_{Al}}$ denotes a sum over all allowed triangulations of
area $A$ and boundary length $l$.


\section{Simulation Methods} \label{sec-methods}
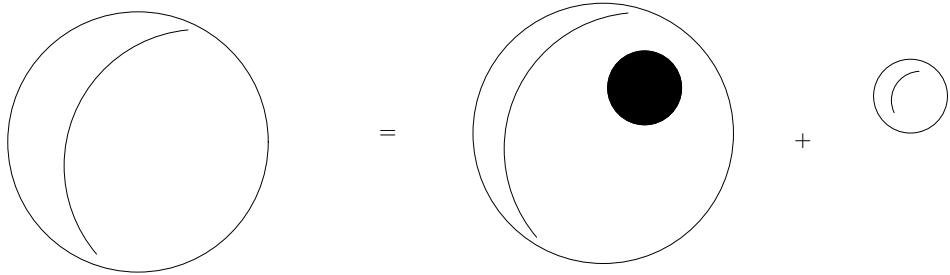
\begin{figure}\begin{center}
    \setlength{\unitlength}{0.00057500in}
\begingroup\makeatletter\ifx\SetFigFont\undefined%
\gdef\SetFigFont#1#2#3#4#5{%
  \reset@font\fontsize{#1}{#2pt}%
  \fontfamily{#3}\fontseries{#4}\fontshape{#5}%
  \selectfont}%
\fi\endgroup%
{\renewcommand{\dashlinestretch}{30}
\begin{picture}(8503,2460)(0,-10)
\put(1750.418,969.422){\arc{2465.984}{2.4383}{4.6186}}
\put(5725.418,1119.422){\arc{2465.984}{2.4383}{4.6186}}
\put(8247.500,1559.500){\arc{525.595}{2.6992}{4.6648}}
\put(1185,1185){\ellipse{2354}{2354}}
\put(5385,1260){\ellipse{2354}{2354}}
\put(8160,1597){\ellipse{670}{670}}
\put(5760,1672){\blacken\ellipse{670}{670}}
\put(5760,1672){\ellipse{670}{670}}
\put(3360,1222){\makebox(0,0)[lb]{\smash{{{\SetFigFont{8}{9.6}{\rmdefault}{\mddefault}{\updefault}=}}}}}
\put(7110,1147){\makebox(0,0)[lb]{\smash{{{\SetFigFont{8}{9.6}{\rmdefault}{\mddefault}{\updefault}+}}}}}
\end{picture}
}
    \caption{Making a disk from a sphere.} \label{sfd}
\end{center}\end{figure}

    A disk can be created from a sphere by simply cutting out a hole,
and this observation forms the basis of our simulation method.  If we
consider a triangulated sphere with a single marked vertex and
imagine cutting out this vertex and all the triangles containing it,
we arrive at a triangulated disk.  Furthermore every triangulated disk
can be reached from an associated triangulated sphere.  This allows us
to sample the disk ensemble by sampling from a sphere ensemble using
the usual procedures \cite{CATTERALL:SOD} --- the only modification
being that we identify one vertex as special, never allowing our local
geometry changing moves to delete it. Additionally we do not endow it
with an Ising spin.  The special vertex along with the ring of
triangles which contain it, form a cap which converts the disk into a
sphere. Fig.~\ref{sfd} shows an illustration.

    Our sampling of triangulations and Ising states is done with a
Metropolis procedure~\cite{MRRTT:EOS}.  Several types of moves are
used in the simulation.  For changes to the geometry, the moves
are link flips, vertex insertions and vertex deletions.  For changes
to the Ising matter, we use either a local spin flip, or a Wolff
cluster algorithm~\cite{WOLFF:CMC}.\footnote{Later revisions of
our code which use versions of these algorithms updated for the
Potts model have been used for some of this work.}

\subsection{Geometry Moves} 
    To change the geometry, we choose a triangle at random, then
choose at random one of 
three types of geometry changing moves:
link flips, vertex insertions, and
vertex deletions.  For vertex insertions, this completely specifies the
move.  For vertex deletions, we need to choose which of the chosen
triangle's three vertices to delete, while for link flips, we need to
choose which of its three links to flip.  Only vertices other than the
special marked vertex and with exactly three surrounding triangles are
allowed to be deleted.  Notice that when a new vertex is added we must
also assign a new Ising spin.  Moves are disallowed if they would lead
to a degenerate triangulation.


    These moves are ergodic in the space of triangulated spheres,
and since every disk can be converted into a sphere, they are also
ergodic in the space of triangulated disks.  In addition to
ergodicity, we require detailed balance.  Let $t$ be a
triangulation, $a$ the set of spins on $t$ unchanged in
the move and $b$ the set of spins on $t$ which
potentially change in the move.  The requirement of detailed balance
is
\begin{eqnarray}
    e^{-S(t, a, b )} T(t, a, b  \rightarrow t',a, b') \\
      = e^{ S(t',a, b')} T(t',a, b' \rightarrow t, a, b ) \nonumber
\end{eqnarray}
where $S(t,a,b)$ is the action and
$T(t,a,b \rightarrow t',a,b')$ is
the transition probability.  Each of our geometry changing updates
requires the transition probabilities satisfy this criteria.  We have
used two types of geometry move in our simulations --- a simple
Metropolis move and a more efficient heat bath style update --- which are
described below.

\subsubsection{Simple Metropolis Algorithm}

    In this method, any spins which will be changed by the move
are chosen at random from a uniform probability distribution in
advance.  In that case we have $b'$ so
\begin{equation}
    T(t,a,b \rightarrow t',a,b') =
    P_p(t,a,b \rightarrow t',a,b') P_a(t,a,b \rightarrow t',a,b')
\end{equation}
where $P_p$ is the probability of proposing the move and $P_a$ is
the probability of accepting the move.  This leads to the
following update probabilities: We accept a proposed vertex
insertion with probability


\begin{equation}
    P_a^i(t,a,b \rightarrow t',a,b') = \frac{A(t)}{A(t')} \mbox{min}\{e^{-S(t',a,b')+S(t,a,b)},1 \}
\end{equation}
where $A(t)$ is the number of triangles in $t$.  A proposed vertex
deletion is accepted with probability


\begin{equation}
    P_a^d(t',a,b'\rightarrow t,a,b) = \frac{1}{2} \mbox{min}\{e^{+S(t',a,b')-S(t,a,b)},1 \}
\end{equation}
    while link flips are accepted with probability
\begin{equation}
    P_a^f(t,a,b \rightarrow t',a,b') = \mbox{min}\{e^{-S(t',a,b')+S(t,a,b)},1 \}.
\end{equation}

\subsubsection{Heat Bath Algorithm}
    In the heat bath variation, insertions or deletions are
accepted or rejected without regard to the state of the vertex's spin.
In the case of insertions, the vertex's spin is chosen from a heat bath
distribution after the move has been accepted.  The probabilities
in this scheme are:

\begin{equation}
    P^d_a(t',a,b' \rightarrow t,a) = \frac{1}{1 + e^{-\Delta S_\uparrow} + e^{-\Delta S_\downarrow}}
\end{equation}
and
\begin{equation}
    P^i_a(t,a \rightarrow t',a,b') = \frac{A(t)}{A(t')}\frac{e^{-\Delta S_\uparrow} + e^{-\Delta S_\downarrow}}
                                                                    {1 + e^{-\Delta S_\uparrow} + e^{-\Delta S_\downarrow}}
\end{equation}
where $\Delta S_\uparrow$ is the change in the action if the
involved spin was or will be chosen to be up, and $\Delta
S_\downarrow$ if it was or will be chosen to be spin down. The
spin of inserted vertices are chosen from:
\begin{equation}
    P(b') = \frac{ e^{-S(t',a,b')} }{ \sum_{b''} e^{-S(t',a,b'')}}.
\end{equation}

\subsection{Spin Updates via Local Spin Flips}
    The procedure for updating a single Ising spin is to choose
one at random and flip it subject to a Metropolis test.  The
probability of accepting a flip is
\begin{equation}
    P_a^s(\sigma \rightarrow -\sigma) = \mbox{min}\{e^{-(S(-\sigma) - S(\sigma))},1\}.
\end{equation}

\subsection{Spin Updates via the Wolff Algorithm}
    The Wolff cluster algorithm \cite{WOLFF:CMC} is an efficient way
of updating Ising (or Potts) spins.  Specifically, it has been shown
to be very efficient at combating the effects of critical slowing near
phase transitions.  The algorithm begins by selecting a vertex at
random. The spin of this vertex is assigned as the first spin of a
cluster. The cluster is grown by examining all the vertex's neighbors
and adding to the cluster any with the same spin as the initial vertex
with probability $1 - e^{-2J}$.  The cluster is then grown further by
following the same procedure with all of the newly added vertices.
The procedure is iterated until no new vertices are added.  Finally,
all the spins in the cluster are reassigned to a (single) random
value. The Wolff cluster algorithm satisfies detailed balance for the
Ising model:
\begin{equation}
    \label{wolff-db}
    \frac{P_p^w(t,a,b \rightarrow t,a,b')}{P_p^w(t,a,b' \rightarrow t,a,b)}
    = \frac{e^{-S_I(t,a,b')}}{e^{-S_I(t,a,b)}}
\end{equation}
where $P_p^w$ is the probability of proposing a move (growing a
given cluster) via the Wolff update.  Since, unlike the Metropolis
algorithm, all moves are accepted this is the same as the
probability of performing the move.  The cluster algorithm relies
on the fact that the action contains only nearest neighbor bond
couplings.  At non-zero boundary field this is not the case and we
have used a generalization of the algorithm which handles non-Ising
like terms.

    Consider an Ising model with such non-Ising like additions to
the action:
\begin{equation}
    S(t,a,b) = S_I(t,a,b) + S_O(t,a,b)
\end{equation}
where $S_I$ is the Ising action and $S_O$ is the new term.  Let
$P_p(t,a,b \rightarrow t,a,b')$ be the probability of proposing a
transition from $(t,a,b)$ to $(t,a,b')$ and $P_a(t,a,b \rightarrow
t,a,b')$ be the probability of accepting the proposal.  Now we require
\begin{equation}
    \label{cluster-db}
    \frac{P_p(t,a,b \rightarrow t,a,b')}{P_p(t,a,b' \rightarrow t,a,b)}\,\frac{P_a(t,a,b \rightarrow t,a,b')}{P_a(t,a,b' \rightarrow t,a,b)}
    =
    \frac{e^{-S_I(t,a,b')}}{e^{-S_I(t,a,b)}}
    \frac{e^{-S_O(t,a,b')}}{e^{-S_O(t,a,b)}}.
\end{equation}

    If we ignore the non-Ising interaction and propose transitions
as the Wolff algorithm prescribes we will have
\begin{equation}
    P_p(t,a,b \rightarrow t,a,b') = P^w_p(t,a,b \rightarrow t,a,b').
\end{equation}
This in conjunction with \ref{wolff-db} and \ref{cluster-db} gives,
\begin{equation}
    \frac{P_a(t,a,b \rightarrow t,a,b')}{P_a(t,a,b' \rightarrow t,a,b)}
    =
    \frac{e^{-S_O(t,a,b')}}{e^{-S_O(t,a,b)}}.
\end{equation}
We satisfy this with the choice
\begin{equation}
    P_a(t,a,b \rightarrow t,a,b') = \mbox{min}\{e^{-(S_O(t,a,b') - S_O(t,a,b))},1\}.
\end{equation}
Thus we modify the usual cluster update so as to subject the entire
move to a Metropolis step dependent {\em only} on the non-Ising like
terms in the action.

    For present purposes, the non-Ising addition to the action is
the interaction of the boundary spins with the boundary magnetic
field:
\begin{equation}
    S_O = -H \sum_{i \in \partial} \sigma_i.
\end{equation}

\subsection{Critical Tuning, Fixing Area and Boundary Length}

    We expect the disk amplitude for the fixed area and boundary
ensemble to be of the form~\cite{TUTTE:ACO,MSS:FLT,COT:AGA}
\begin{equation}
    \label{eq-fixed-amp}
    Z(l,A) = l^a e^{\nu_c l} A^b e^{\mu_c A} e^{-\frac{l^2}{A}}.
\end{equation}
In the above, $\mu_c$ and $\nu_c$ are the critical values of the cosmological
constants.

Given the simulation methods we have described, both area $A$ and
boundary length $l$ will fluctuate. We can not prevent these fluctuations
without sacrificing ergodicity.  However, if we modify the action
by adding a geometry fixing part
\begin{equation}
    S_f(\lambda_\nu,\lambda_\mu,l_0,A_0)
    = \lambda_\nu (l - l_0)^2 + \lambda_\mu (A - A_0)^2
\end{equation}
we can force the simulation to remain in the vicinity of some target
area $A_0$ and boundary length $l_0$.  Notice that
these modification terms vanish for disks with these target values and
we can hope to recover the results of the fixed area/boundary length
ensemble with $A=A_0$ and $l=l_0$ by merely sampling such target disks
from our ensemble.  

With the above addition, the simulation will visit states with weight
\begin{eqnarray}
    W(l,A) &=& e^{-S_f(\lambda_\nu,\lambda_\mu,l_0,A_0)}Z_{lA} \\
           &=& l^a e^{(\nu_c - \nu)l - \lambda_\nu (l - l_0)^2} A^b e^{(\mu_c - \mu)A -\lambda_\mu (A - A_0)^2} e^{-\frac{l^2}{A}} \nonumber
\end{eqnarray}
If we assume that the average values of $l$ and $A$ will occur
approximately where they maximize $W$, we have
\begin{eqnarray}
    \label{maxweightA}
    \left. \frac{\partial}{\partial A} W(A,l) \right|_{\left< l \right>,\left< A \right>} &=& 0 \\
    \label{maxweightl}
    \left. \frac{\partial}{\partial l} W(A,l) \right|_{\left< l \right>,\left< A \right>} &=& 0
\end{eqnarray}
which we can solve to get
\begin{equation}
    \mu_c = -\frac{\left<l\right>^2}{\left<A\right>^2} - \frac{b}{\left<A\right>} + \mu + 2 \lambda_\mu (\left<A\right> - A_0)
\end{equation}
and
\begin{equation}
    \nu_c = -\frac{a}{\left<l\right>} + \nu +
    2 \lambda_\nu (\left<l\right> - l_0) - \frac{2\left<l\right>}{\left<A\right>}.
\end{equation}
These equations allow us, as a by-product, to read off estimates for
the critical couplings $\mu_c$ and $\nu_c$ by measuring the mean
area $\left<A\right>$ and boundary length $\left<l\right>$ . Of course
to do this, in general, we need the values of the exponents $a$ and
$b$ --- numbers that are not generally known (see Section~\ref{sec-da} for a
counter example). Notice, however, that the values of $a$ and $b$ are
irrelevant in the thermodynamic limit ($A_0,l_0 \rightarrow \infty$)
so long as $l_0 \ll A_0$.  This condition will be met close to the
tricritical point where $l \sim \sqrt{A}$.

\section{Disk Amplitude and Strong Coupling} \label{sec-tests}

    We have tested our simulation methods in two ways.  In the first,
which is akin to the small coupling expansion, we measure the
frequency of visits to small triangulations and compare the results to
known values.  In the second, we measure pure gravity disk amplitudes
and compare estimates of the exponents for area and boundary length
with known values.

\subsection{Strong Coupling}
    We may recast the partition function in the following form
\begin{equation}
    Z = \left( \omega_1 e^{-\mu} + \omega_2 e^{-2\mu } + \omega_3 e^{-3\mu} \ldots \right)
\end{equation}
where $\omega_A$ represents the weight associated with triangulations
of area $A$ with all other variables summed over. For large
$\mu$, only the smallest triangulations survive and the coefficients
$\omega_A$ may be calculated explicitly.  This is referred to as the
strong coupling expansion.  Computationally, we can achieve the same
result by explicitly limiting the simulation to triangulations less
than a certain area.  Unlike the above, this procedure does not
require large $\mu$.

    We hand counted the number of distinct triangulations for disks
with areas up to five.  The counting is weighted by inverse symmetry
factors, the calculations of which were the most difficult part of the
counting.  For the case of $\mu = \nu = 0$ and no Ising spins, the
number of times the simulation visits triangulations of a given area
and boundary length should be proportional to this count.  In
Tab.~\ref{AFSD} we compare our hand calculated counts to the observed
number of visitations by the simulation.  The data are normalized so
that the values in the first row are one.  The agreement was excellent
and served as an important confirmation of the correctness of our
code.

\begin{table}
    \begin{center}
        \begin{tabular}{|r|r|r|} \hline
            $A,l$ & count & visits \\ \hline
            $1,3$ &  $1.00$ &  $1.00 \pm 0.00 $ \\ \hline
            $2,4$ &  $1.50$ &  $1.50 \pm 0.01 $ \\ \hline
            $3,3$ &  $1.00$ &  $0.99 \pm 0.01 $ \\ \hline
            $3,5$ &  $3.00$ &  $2.99 \pm 0.01 $ \\ \hline
            $4,4$ &  $3.75$ &  $3.75 \pm 0.03 $ \\ \hline
            $4,6$ &  $7.00$ &  $6.96 \pm 0.10 $ \\ \hline
            $5,3$ &  $3.00$ &  $2.99 \pm 0.01 $ \\ \hline
            $5,5$ & $12.60$ & $12.60 \pm 0.10 $ \\ \hline
            $5,7$ & $18.00$ & $17.93 \pm 0.20 $ \\ \hline
        \end{tabular}
    \end{center}
    \caption{Comparison of the number of distinct triangulations for a given area and boundary to the frequency of visitations by the simulation with no Ising spins.}
    \label{AFSD}
\end{table}

    We extended this calculation to triangles with Ising spins and
possible boundary fields. The results, shown in Tabs.~\ref{AFSDWISON1}~and~\ref{AFSDWISON2},  were also excellent.

\begin{table}
    \begin{center}
        \begin{tabular}{|r|r|r|} \hline
            $A,l$ &  count     & visits \\ \hline
            $1,3$ & $   1.00 $ & $    1.00 \pm   0.00 $\\ \hline
            $2,4$ & $   3.03 $ & $    3.03 \pm   0.04 $\\ \hline
            $3,3$ & $   2.04 $ & $    2.03 \pm   0.04 $\\ \hline
            $3,5$ & $  12.27 $ & $   12.24 \pm   0.30 $\\ \hline
            $4,4$ & $  15.44 $ & $   15.43 \pm   0.30 $\\ \hline
            $4,6$ & $  57.89 $ & $   57.78 \pm   1.00 $\\ \hline
            $5,3$ & $  12.45 $ & $   12.45 \pm   0.30 $\\ \hline
            $5,5$ & $ 104.94 $ & $  104.73 \pm   3.00 $\\ \hline
            $5,7$ & $ 301.05 $ & $  300.61 \pm   5.00 $\\ \hline
        \end{tabular}
    \end{center}
    \caption{Comparison of the number of distinct triangulations for a given area and boundary to the frequency of visitations by the simulation with Ising spins included with $J = 0.1$ and $H = 0$.}
    \label{AFSDWISON1}
\end{table}

\begin{table}
    \begin{center}
        \begin{tabular}{|r|r|r|} \hline
            $A,l$ &  count     & visits \\ \hline
            $1,3$ & $   1.00 $ & $    1.00 \pm   0.00 $\\ \hline
            $2,4$ & $   5.29 $ & $    5.30 \pm   0.20 $\\ \hline
            $3,3$ & $   2.07 $ & $    2.09 \pm   0.10 $\\ \hline
            $3,5$ & $  37.38 $ & $   37.44 \pm   2.00 $\\ \hline
            $4,4$ & $  27.17 $ & $   27.31 \pm   1.00 $\\ \hline
            $4,6$ & $ 308.57 $ & $  309.93 \pm  20.00 $\\ \hline
            $5,3$ & $  12.76 $ & $   12.81 \pm   1.00 $\\ \hline
            $5,5$ & $ 320.80 $ & $  321.67 \pm  20.00 $\\ \hline
            $5,7$ & $2808.60 $ & $ 2820.27 \pm 100.00 $\\ \hline
        \end{tabular}
    \end{center}
    \caption{Comparison of the number of distinct triangulations for a given area and boundary to the frequency of visitations by the simulation with Ising spins, a Boundary Magnetic Field with $J = 0.1$ and $H = 1$.}
    \label{AFSDWISON2}
\end{table}


\subsection{Disk Amplitude}
\label{sec-da}
    If we eliminate the Ising spins and tune the cosmological
constants to their critical values, we obtain the disk amplitude which
is predicted to take the form~\cite{TUTTE:ACO,MSS:FLT,COT:AGA}
\begin{equation}
    Z(A,l) = l^{a}A^{b}e^{-\frac{l^2}{A}}.
\end{equation}
The exponents $a$ and $b$ are predicted to be $-\frac{1}{2}$ and
$-\frac{5}{2}$ respectively.  We have attempted to measure these
exponents from our simulations.  The exponent $b$ can be found by
measuring the fraction of bulk vertices which are 3-fold coordinated,
while $a$ can be obtained by fitting the so-called `baby universe'
distribution.

\subsubsection{Measuring the exponent $b$}
    Let $t_1$ be a triangulation os a disk of area $A - 2$ and boundary length $l$.
Let $t_2$ and $t_3$ be disks of area $A$ and boundary length $l$.
Detailed balance tells us
\begin{equation}
    W(t_1) P(t_1 \rightarrow t_2) = W(t_2) P(t_2 \rightarrow t_1)
\end{equation}
where $W(t_1)$ is the weight of state $t_1$ and $P(t_1 \rightarrow t_2)$ is
the probability of a transition from $t_1$ to $t_2$.  From this we get
\begin{equation}
    W(t_1) \sum_{t_2} P(t_1 \rightarrow t_2) = \sum_{t_2} W(t_2) P(t_2 \rightarrow t_1).
\end{equation}
We can use this to write the partition function as
\begin{eqnarray}
    Z(A-2,l) &=& \sum_{t_1} W(t_1) \\
             &=& \sum_{t_2} W(t_2) \sum_{t_1} \frac{P(t_2 \rightarrow t_1) }{ \sum_{t_3} P(t_1 \rightarrow t_3)}.
\end{eqnarray}
Now consider the ratio
\begin{eqnarray}
    \frac{Z(A-2,l)}{Z(A,l)}
    &=& \left< \sum_{t_1} \frac{P(t_2 \rightarrow t_1) }{ \sum_{t_3} P(t_1 \rightarrow t_3)} \right>_{W(t_2)}.
\end{eqnarray}
Since we have eliminated the Ising spins, the action is $0$ for all
states and the transition probabilities are $1$ for all possible
moves.  That being the case, the sums over transition probabilities
just count the number of possible moves.  The number of possible insertions is equal
to the number of triangles into which a new vertex can be inserted:
\begin{equation}
    \sum_{t_3} P(t_1 \rightarrow t_3) = A-2,
\end{equation}
while the number of possible deletions is equal to the number of bulk vertices with exactly three
surrounding triangles, $N_3$,: 
\begin{equation}
    \sum_{t_1} P(t_2 \rightarrow t_1) = N_3.
\end{equation}
So,
\begin{equation}
    \frac{Z(A-2,l)}{Z(A,l)} = \left< \frac{N_3}{A-2} \right>_{W(t_2)}.
\end{equation}
Given the expected form of $Z$,
\begin{eqnarray}
    \left<\frac{N_3}{A-2}\right>_{W(t_2)} &=& \left(\frac{A-2}{A}\right)^b e^{-\frac{l^2}{A-2} + \frac{l^2}{A}}\\
    \ln \left<\frac{N_3}{A-2}\right>_{W(t_2)}
    &=& -\frac{2b}{A} +O\left( \frac{1}{A^2} \right)
\end{eqnarray}

\begin{figure}\begin{center}
    \includegraphics[height=7cm]{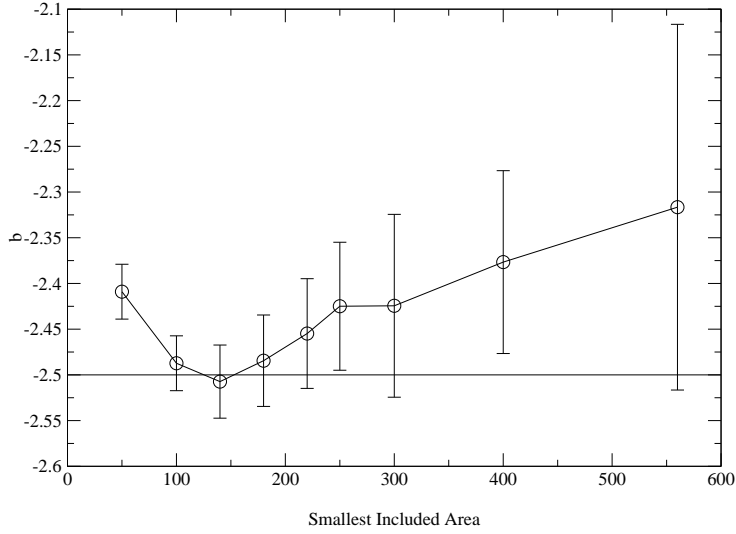}
    \caption{$b$ vs. smallest included volume.}
    \label{bvsv}
\end{center}\end{figure}

    We generated data with areas between 50 and 1600 to test this
and found it was statistically consistent with $b = -2.5$. Because the
above partition function is only correct for large volumes, the best
results are obtained by excluding disks smaller than some cutoff from
the fits.  Of course, excluding too much data will harm the fits ---
however our data shows a robust plateau close to the predicted value.
This can be seen in Fig.~\ref{bvsv}.

\subsubsection{Baby Universes and the exponent $a$}
    When two sections of the disk share a boundary of only one
link, we say that the disk has split into two minimal neck baby
universes.  Using the form of the partition function mentioned
above, the probability of a disk of area $A$ and boundary $l$
generating a baby universe of area $A_1$ and boundary $l_1$ is
\begin{eqnarray}
    P[A_1,l_1;A,l] &=& \frac{Z(A_1,l_1) Z(A - A_1,l + 2 - l_1)}{Z(A,l)}\\
                   &=& \left(\frac{l_1(l+2-l_1)}{l}\right)^a
                       \left(\frac{A_1(A-A_1)}{A}\right)^b \\ \nonumber
                           && \exp\{-\frac{l_1^2}{A_1^2} - \frac{(l+2-l_1)^2}{(A-A_1)^2} - \frac{l^2}{A^2}\}
\end{eqnarray}
We generated histograms of baby universe observations with various
fixed disk areas and boundary lengths.  To do this we found all such
minimal necks on various equilibrium configurations and accumulated
2-d histograms labelled by the areas and boundary lengths of the
smaller baby universes.  Using the analytical value of $b = -2.5$, we
found the value of $a$ which gave the best fit of the above
probability distribution to the data.  These are shown in
Tab~\ref{AFIT}.  The result for $A = 2000$ is within $2\sigma$ of the
predicted value $a = -\frac{1}{2}$ and we believe that the
discrepancies for $A = 1000$ and $A = 500$ are likely finite size
effects.

\begin{table}
    \begin{center}
        \begin{tabular}{|r|r|r|r|} \hline
            $A,l$      & $a$   & $\sigma$ & $\chi_{\rm d.o.f.}^2$ \\ \hline
             500,  32 & -0.451 & 0.0008 & 246.2 \\ \hline
            1000,  44 & -0.487 & 0.001  &  41.0 \\ \hline
            2000,  64 & -0.498 & 0.001  &  13.4 \\ \hline
        \end{tabular}
    \end{center}
    \caption{$a$ which gives best fit to baby universe histogram.}
    \label{AFIT}
\end{table}

\section{Phase Diagram with $H=0$} \label{sec-phases}
For the work described in this section, both cosmological constants
were set to their critical values.  Areas were kept fixed, but
$\lambda_\nu$ was set to zero, allowing the boundary length to
fluctuate.  We have identified phases based on two observables,
magnetization and boundary length.  To locate phase transitions we
look for peaks in the corresponding susceptibilities.  We define
the magnetization of a given disk to be
\begin{equation}
    M = \frac{1}{N} \left|\sum_i \sigma_i\right|
\end{equation}
where $N$ is the number of vertices in the disk, and the sum is
over all vertices.  We use angle brackets to indicate averages
over sampled disks, so $\left<M\right>$ refers to an average of
$M$ over disks generated by the simulation.  The magnetic
susceptibility is
\begin{equation}
    \chi_M = \frac{\partial \left<M\right>}{\partial J} = \left< N M M \right> - \left<N M\right> \left<M\right>.
\end{equation}
When we refer to boundary susceptibility, we mean
\begin{equation}
    \chi_L = \frac{\partial \left<L\right>}{\partial \nu} = \left<L^2\right> - \left<L\right>^2.
\end{equation}

\begin{figure}\begin{center}
    \includegraphics[height=7cm]{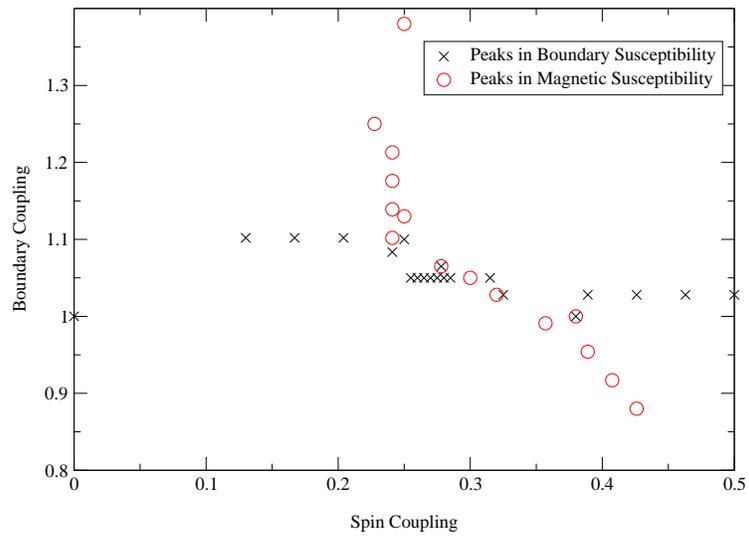}
    \caption{Observed peaks in susceptibilities for $A = 1000$, $H = 0$.}
    \label{phases}
\end{center}\end{figure}

\begin{figure}\begin{center}
    \includegraphics[height=7cm,height=8cm]{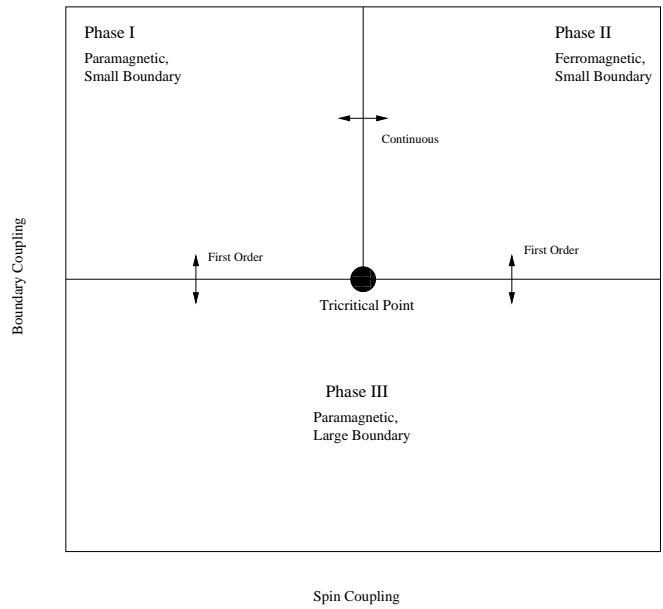}
    \caption{Schematic phase diagram at $H = 0$.}
    \label{phase_schematic}
\end{center}\end{figure}

    By these criteria we have found three phases --- a paramagnetic
phase with a small boundary (I), a ferromagnetic phase with a small
boundary (II) and a paramagnetic phase with a large boundary (III).
The locus of susceptibility peaks for $A = 1000$ is shown in
Fig.~\ref{phases} and a schematic phase diagram is shown in
Fig.~\ref{phase_schematic}

    We used finite size scaling to determine the order of the observed
transitions.  To make clear possible finite size effects, we looked at
exponents for successive pairs of volumes.  The scaling exponent for
some quantity $x$ with values $x_1$ at area $A_1$ and $x_2$ at area
$A_2$ is given by
\begin{equation}
    \frac{\ln(x_2/x_1)}{\ln(A_2/A_1)}.
\end{equation}

\begin{figure}\begin{center}
    \includegraphics[height=7cm]{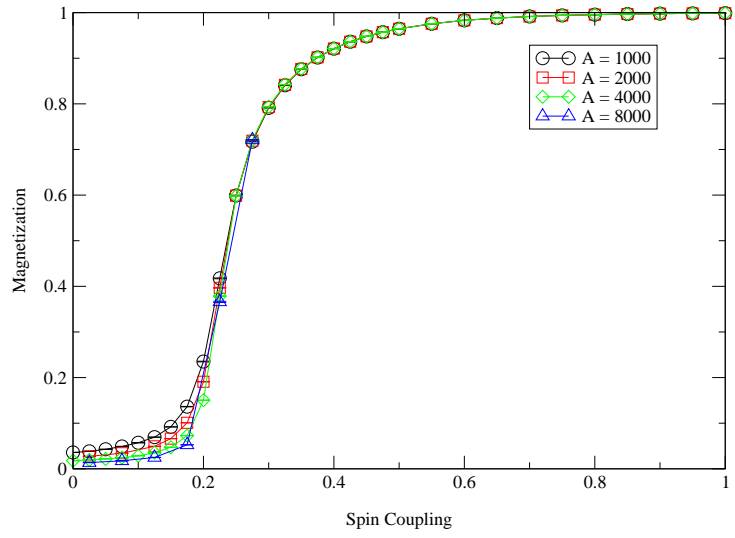}
    \caption{Magnetization, $\nu = 5$, $H = 0.$}
    \label{m_kb5}
\end{center}\end{figure}

\begin{figure}\begin{center}
    \includegraphics[height=7cm]{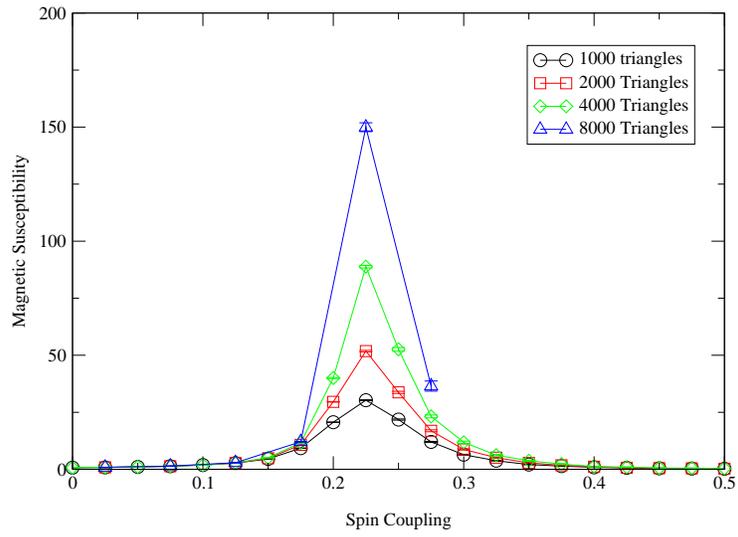}
    \caption{Magnetic susceptibility, $\nu = 5$, $H = 0$.}
    \label{chi_m_kb5}
\end{center}\end{figure}

   For sufficiently large boundary cosmological constant, $\nu$, the boundary length is minimal and we
expect behavior similar to that of a marked sphere.  In that case,
there should be a continuous magnetic phase transition at $J_c=
-\frac{1}{2} \ln \tanh \left( \frac{1}{2} \ln \frac{108}{23} \right)
\approx 0.216273$.  The observed magnetization and susceptibility in
Fig.s~\ref{m_kb5}~and~\ref{chi_m_kb5} show this.  Continuum
calculations for 2-d quantum gravity coupled to $c = \frac{1}{3}$
matter predict $\left<N M^2\right> \sim A^{\frac{2}{3}}$
\cite{BFHM:TIM}.  Tab.~\ref{M_SCALING} shows values of this exponent
extracted from data at succeeding pairs of areas.  The agreement is
quite good for the larger areas.

\begin{table}
    \begin{center}
        \begin{tabular}{|l|l|} \hline
            Areas           & $\ln(x_2/x_1)/\ln(A_2/A_1)$ \\ \hline
             $1000 -  2000$ & $0.70 \pm 0.01$ \\ \hline
             $2000 -  4000$ & $0.70 \pm 0.01$ \\ \hline
             $4000 -  8000$ & $0.69 \pm 0.02$ \\ \hline
             $8000 - 16000$ & $0.68 \pm 0.02$ \\ \hline
            $16000 - 32000$ & $0.67 \pm 0.04$ \\ \hline
        \end{tabular}
     \end{center}
    \caption{Scaling exponents of $x = \left<N M^2\right>$ at transition between phases I and II.}
    \label{M_SCALING}
\end{table}

\begin{figure}\begin{center}
    \includegraphics[height=7cm]{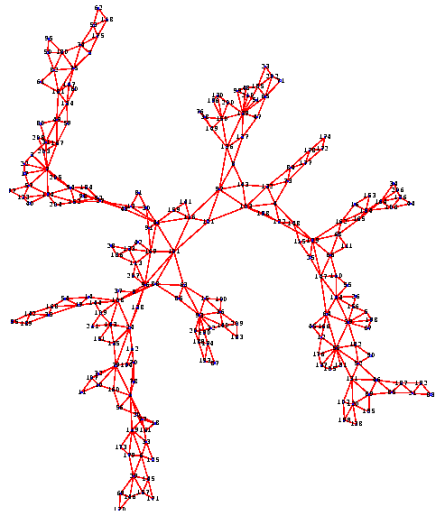}
    \caption{A triangulation, $J = 5, \nu = -5$, $H = 0$.}
    \label{triangulation}
\end{center}\end{figure}

\begin{figure}\begin{center}
    \includegraphics[height=7cm]{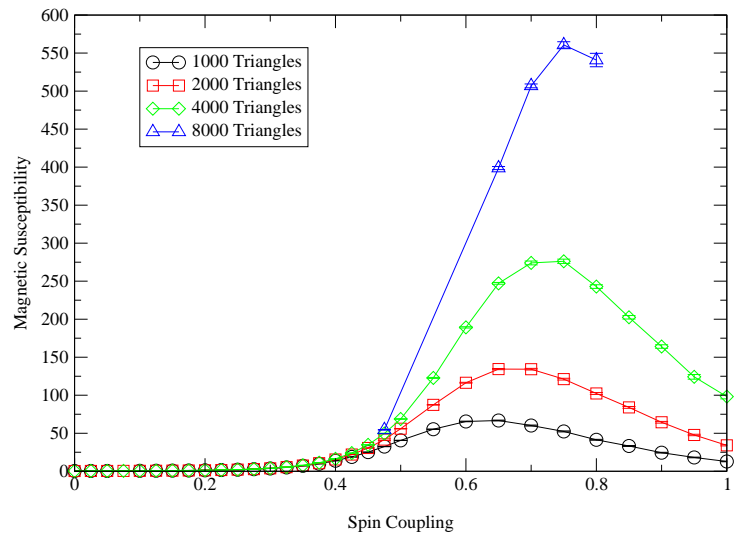}
    \caption{Magnetic susceptibility, $\nu = -5$, $H = 0$.}
    \label{chi_m_kb-5}
\end{center}\end{figure}

\begin{figure}\begin{center}
    \includegraphics[height=7cm]{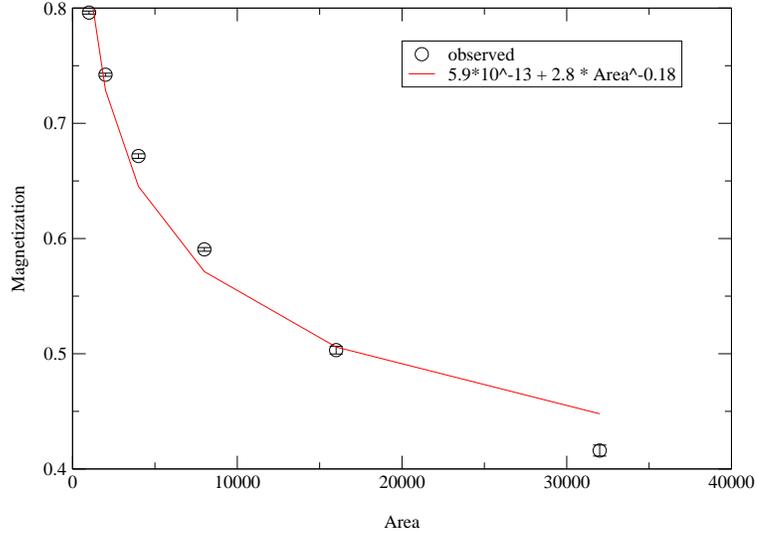}
    \caption{Magnetization vs. area, $J = 0.75, \nu = -5$, $H = 0$.}
    \label{m_vs_a_j0.75_kb-5}
\end{center}\end{figure}

    For sufficiently small $\nu$, the boundary length is maximal and,
as shown in Fig.~\ref{triangulation}, the geometry is like that of
a branched polymer.  In this essentially 1-d configuration we expect
that the system will not magnetize.  However, from
Fig.s~\ref{phases}~and~\ref{chi_m_kb-5}, it appears that the line
between Phases I and II continues into small $\nu$.  Notice, though,
that the susceptibility scaling exponents on this continuation, shown
in Tab.~\ref{M_SCALING_2}, decrease as the disk area is increased.
This indicates that the continuation is a finite size artifact.  To
confirm this we examined how the mean magnetization varies with area
at $\nu = -5$ and $J = 0.75$ which would be within this
small-boundary, ferromagnetic phase if it existed.  We fit the data to
the form $c_0 + c_1 * A ^{c_2}$ with the constraints $0 <= c_0, c_1 <=
1$.  As shown in Fig.~\ref{m_vs_a_j0.75_kb-5}, extrapolation based
on this fit implies that the magnetization will approach zero as the
area is sent to infinity.  To the extent that the fit does not agree
with the data, the fit seems to be underestimating the rate of
decrease so the predication appears safe.  Our conclusion is that for
small $\nu$ there is only one phase (III) characterized by a large
boundary and unmagnetized spins.

\begin{table}
    \begin{center}
        \begin{tabular}{|l|l|} \hline
            Areas   & $\ln(x_2/x_1)/\ln(A_2/A_1)$ \\ \hline
             $1000 -  2000$ & $0.832 \pm 0.006 $ \\ \hline
             $2000 -  4000$ & $0.765 \pm 0.008 $ \\ \hline
             $4000 -  8000$ & $0.68  \pm 0.01  $ \\ \hline
             $8000 - 16000$ & $0.60  \pm 0.02  $ \\ \hline
            $16000 - 32000$ & $0.51  \pm 0.04  $ \\ \hline
        \end{tabular}
    \end{center}
    \caption{Scaling exponents of $x = \left<N M^2\right>$ for $\nu = -5$ and $J = 0.75$, $H = 0$.}
    \label{M_SCALING_2}
\end{table}

%

\begin{figure}\begin{center}
    \includegraphics[height=7cm]{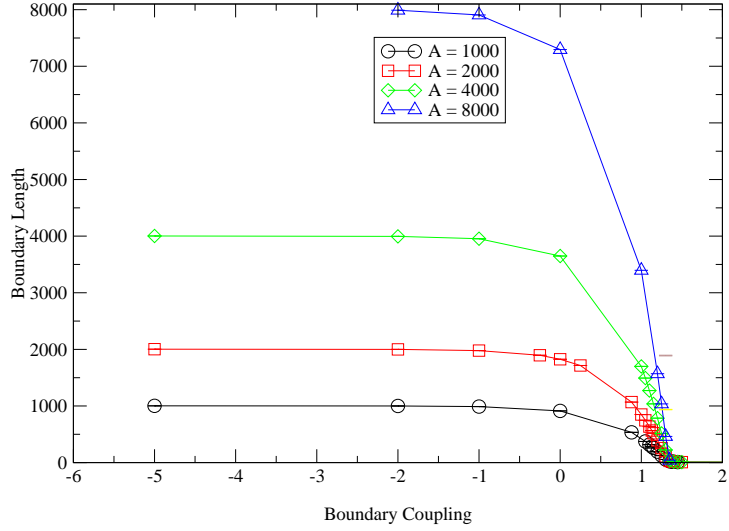}
    \caption{Boundary length, $J = 1$, $H = 0$.}
    \label{l_j1}
\end{center}\end{figure}

Fig.~\ref{l_j1} and Tab.~\ref{L_SCALING} show that along the
transition from small (I,II) to large (III) boundary phases, the mean
boundary length appears to jump discontinuously as $\nu$ is varied and
the boundary length susceptibility exhibits a critical exponent of
approximately one.  This is strong evidence that the transition is
first order. 

\begin{table}
    \begin{center}
        \begin{tabular}{|l|l|} \hline
            Areas        & $\ln(x_2/x_1)/\ln(A_2/A_1)$ \\ \hline
             $1000 -  2000$ & $1.38 \pm 0.01 $\\ \hline
             $2000 -  4000$ & $1.07 \pm 0.02 $\\ \hline
             $4000 -  8000$ & $0.97 \pm 0.02 $\\ \hline
	     $8000 - 16000$ & $0.99 \pm 0.002$ \\ \hline
	    $16000 - 32000$ & $0.96 \pm 0.02 $\\ \hline
        \end{tabular}
    \end{center}
    \caption{Scaling exponent of $x = \left<l^2\right> - \left<l\right>^2$ for $J = 1.0$, $\nu = 1.3$ and $H = 0$.}
    \label{L_SCALING}
\end{table}


\section{Scaling Behavior with a Boundary Field} \label{sec-boundary}

    At the tricritical point, we performed simulations to test the
predictions of \cite{COT:BFA} regarding the scaling of the bulk
magnetization, the boundary magnetization and their susceptibilities
with the boundary magnetic field.  For this part of the
investigation, the Ising coupling was set to its continuum
critical value.  To reduce the number of parameters we needed to
keep track of, the boundary lengths were fixed at $\sqrt{A}$.
While this does involve tuning $\nu$ to its critical value,
the boundary-length-fixing Gaussian in the action makes its exact
value unimportant.  Had we not done this, the approximate nature
of the boundary coupling tuning procedure would have made
comparing data from different runs more difficult.

    The existence of a boundary magnetic field breaks the symmetry
between up and down spins, so we no longer need to take the modulus of
the spin sums.  We can now define magnetizations
\begin{equation}
    M_{\not\partial} = \frac{1}{N - l} \sum_{i \notin \partial} \sigma_i
\end{equation}
\begin{equation}
    M_{\partial} = \frac{1}{l} \sum_{i \in \partial} \sigma_i.
\end{equation}
The susceptibilities are defined in the usual manner.
Expressions for both boundary and bulk magnetizations are derived
in \cite{COT:BFA}:
\begin{equation}
    \left<M_{\partial}\right> = \frac{(e^{2H}-1)(3 + (2+\sqrt{7})e^{2H})}
                                            {(1 + (-1 + \sqrt{7})e^{2H} + (2 + \sqrt{7}e^{4H})}
\end{equation}
and
\begin{equation}
    \label{M_bulk}
    \left<M_{\not\partial}\right> \sim L^{\frac{1}{3}} A^{-\frac{1}{3}}
\end{equation}
where
\begin{equation}
    \label{alpha}
    L = \alpha(H) l
\end{equation}
and
\begin{equation}
    \label{alpha-eq}
    \alpha(H) = \frac{e^{2H} (1 + e^{2H})}
                     {1 + (-1 + \sqrt{7})e^{2H} + (2 + \sqrt{7}) e^{4H}}.
\end{equation}

\begin{figure}\begin{center}
    \includegraphics[height=7cm]{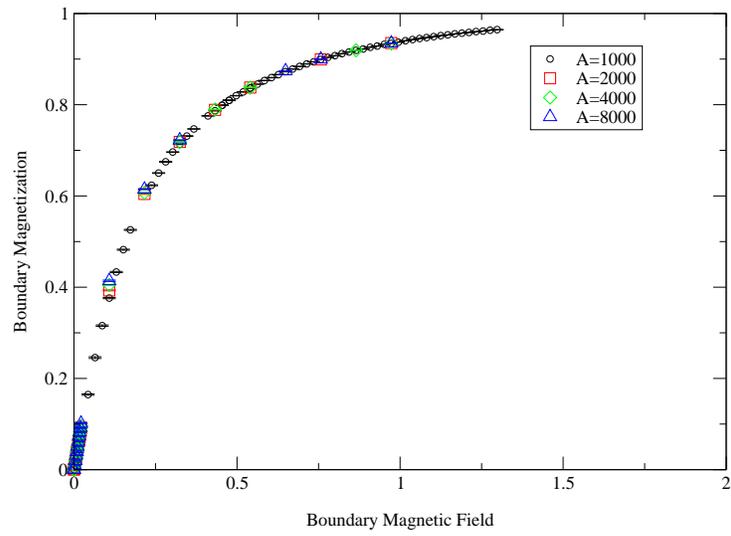}
    \caption{Boundary magnetization vs. boundary magnetic field at the tricritical point.}
    \label{bom_scaling}
\end{center}\end{figure}

\begin{figure}\begin{center}
    \includegraphics[height=7cm]{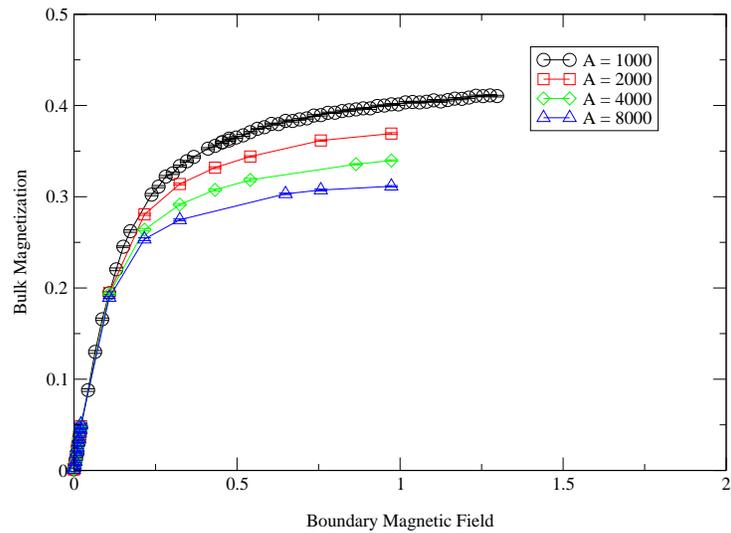}
    \caption{Bulk magnetization vs. boundary magnetic field at the tricritical point.}
    \label{bum_scaling}
\end{center}\end{figure}

However, at least some of these results are dependent on the exact
nature of the discretization procedure, i.e. the nature of the lattice
ensemble.  Our ensemble of so-called {\it combinatorial}
triangulations is a subset of the set of triangulations employed by
these authors. In concrete terms we might expect that the function
$\alpha(H)$ relevant to a description of the combinatorial ensemble
might differ substantially from the function given above. In spite of
this it is reasonable to expect qualitatively similar behavior in the
two cases. We have therefore looked for a boundary magnetization which
increases with $H$ and is independent of $A$ and $l$, as well as a
bulk magnetization which decreases with $H$ for fixed $l$.  The
boundary magnetization plot shown in Fig.~\ref{bom_scaling} indeed
illustrates this predicted behavior.  In contrast,
Fig.~\ref{bum_scaling} shows the bulk magnetization increases with
$H$.

\begin{figure}\begin{center}
    \includegraphics[height=7cm]{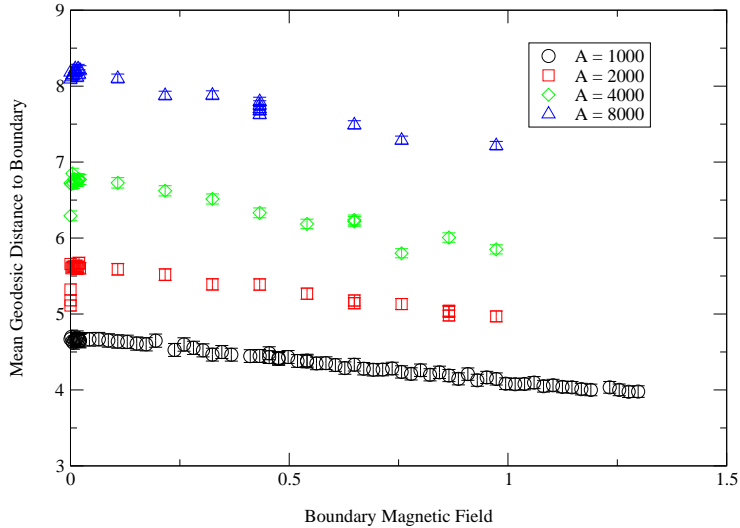}
    \caption{Mean geodesic distance from boundary vs. boundary magnetic field at the tricritical point.}
    \label{mean_dist_from_boundary}
\end{center}\end{figure}

    Caroll, Ortiz and Taylor speculate that their predicted decrease
of bulk magnetization with $H$ is due to an increase in the average
spin's distance from the boundary when the boundary field is
increased.  To test this, we measured geodesic distances from all
vertices on the disks to the boundary as follows.  All boundary
vertices are labelled as having a geodesic distance from the boundary
of zero.  All vertices connected to the boundary vertices and not
already labelled are labelled as having a geodesic distance from the
boundary of one.  We continue labelling successively deeper vertices
until there are none left.  At that point we compute the average
distance from the boundary.  We observed behavior opposite that
predicted by Caroll, Ortiz and Taylor: the mean distance to the
boundary, illustrated in Fig.~\ref{mean_dist_from_boundary} {\it
decreases} with increasing magnetic field.  This behavior is then
consistent with the behavior of the bulk magnetization.

    An additional prediction is that the boundary magnetic
susceptibility should be finite, $\chi_{\partial} = \frac{1 + 2
\sqrt{7}}{3}$, in the limit of $H \rightarrow 0$ and $A \rightarrow
\infty$.  This differs markedly from the flat-space behavior of
infinite susceptibility \cite{MW:TTDIM} which exhibits a logarithmic
singularity. To compare this result with the numerical simulations we
must be careful. The matrix model result is a prediction for the
behavior of the boundary susceptibility on an infinite area disk as
the boundary magnetic field is tuned to zero.  Our simulations are
carried out at finite area and so we should not take $H$ all the way
to zero to do the comparison (the magnetic field effectively fails to
break the symmetry of the Ising action for fields $H<1/l$). Thus we
are led to examine the dependence of the susceptibility on the disk
area for small but non-zero magnetic fields.  In
Tab.~\ref{bound_chi_scale} we show finite size scaling exponents for
the boundary magnetic susceptibility at three small boundary magnetic
field strengths. While the exponents at $H = 0.0216$ are positive,
which would indicate a divergence, this field strength is below the
value needed to break the spin-up, spin-down symmetry for the smallest
disks of area $1000$ ($\frac{1}{\sqrt{1000}} \approx 0.0316$). For the
larger values of $H$, we find negative exponents which imply a finite
value of the susceptibility for small magnetic field in the infinite
area limit.  This is in qualitative agreement with the matrix model
predictions and quite different from the flat space result.

\begin{table}
    \begin{center}
        \begin{tabular} {|r|r|r|} \hline
            $H$    & $A1, A2$  & $\frac{\log{\chi_2 / \chi_1}}{\log{l_2 / l_1}}$ \\ \hline
            0.0216 & 1000,2000 & 0.0863 \\
                   & 2000,4000 & 0.0891 \\
                   & 4000,8000 & 0.0651 \\ \hline
            0.108  & 1000,2000 & -0.0190 \\
                   & 2000,4000 & -0.0218 \\
                   & 4000,8000 & -0.0197 \\ \hline
            0.216  & 1000,2000 & -0.0490 \\
                   & 2000,4000 & -0.0361 \\
                   & 4000,8000 & -0.0636 \\ \hline
        \end{tabular}
    \end{center}
    \caption{Critical exponent of boundary magnetic susceptibility at the tricritical point.}
    \label{bound_chi_scale}
\end{table}

\begin{figure}\begin{center}
    \includegraphics[height=7cm]{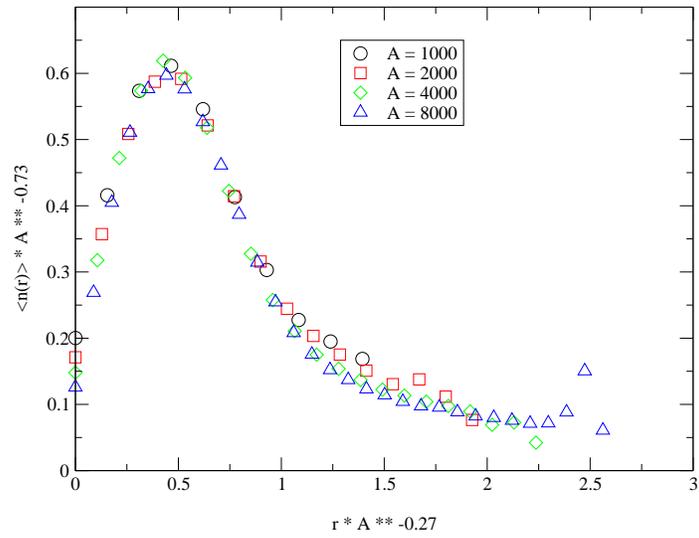}
    \caption{Number of vertices at geodesic distance $r$ from the boundary at the tricritical point with $H = 0.433$.}
    \label{geo_hist}
\end{center}\end{figure}

\begin{figure}\begin{center}
    \includegraphics[height=7cm]{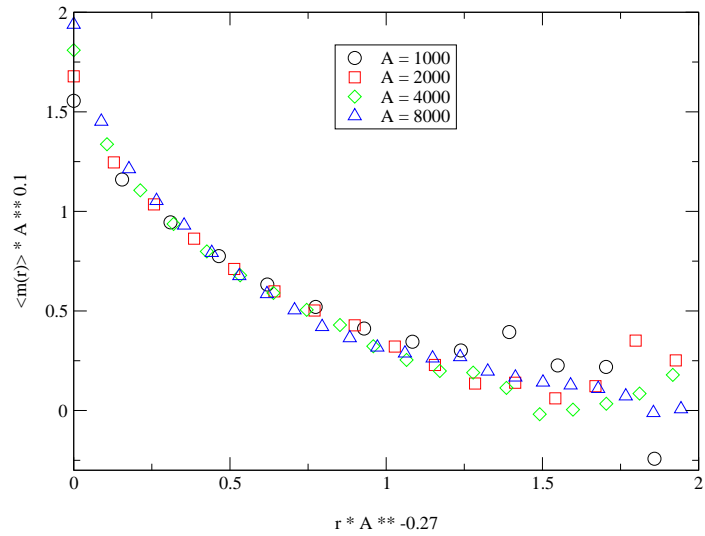}
    \caption{Magnetization of vertices at geodesic distance $r$ from the boundary at the tricritical point with $H = 0.433$.}
    \label{geo_mag}
\end{center}\end{figure}

\begin{figure}\begin{center}
    \includegraphics[height=7cm]{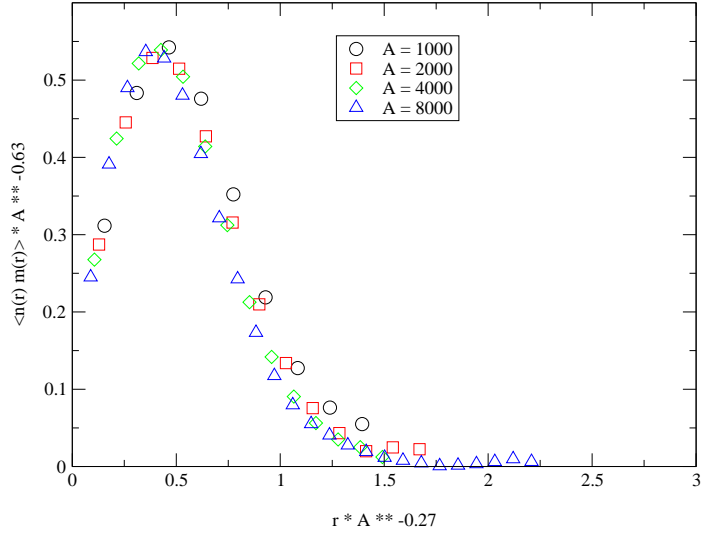}
    \caption{Sum of spins of vertices at geodesic distance $r$ from the boundary at the tricritical point with  $H = 0.433$.}
    \label{geo_spin_sum}
\end{center}\end{figure}

We have looked at the behavior of three geodesic correlation functions
which measure the average properties of the spins as a function of
distance to the boundary.  The first of these, $n_1(r,N)$, measures
the mean number of vertices encountered at distance $r$ from the
boundary.  The second, $n_\sigma(r;N)$, measures the net (integrated)
magnetization and the third, $M(r;N)$, the mean magnetization at
distance $r$ from the boundary.  They are defined explicitly as follows
 \begin{eqnarray}
    n_1(r;N) =        & \left< \sum_i \delta_{D_i,r} \right>                                         &\sim N^{1-1/d_H} F_1(r/N^{1/d_H}) \\
    n_{\sigma}(r;N) = & \left< \sum_i \sigma_i \delta_{D_i,r} \right>                                & \sim N^{1-\Delta-1/d_H} F_{\sigma}(r/N^{1/d_H}) \\
    M(r;N) =          & \left< \sum_i \frac{\sigma_i \delta_{D_i,r}}{\sum_k \delta_{D_k,r}} \right>  & \sim N^{-\Delta} g_{M}(r/N^{1/d_H})
\end{eqnarray}
where we have also introduced finite size scaling forms for these
three functions analogous to the functions used for geodesic
correlation functions defined on the sphere in \cite{AA:QG}. Requiring
that the area ($N \sim A$) dependence in $M(r;N)$ disappear for small
$r$ implies that this correlation function takes on a simple power law
form for small $x$
\begin{equation}
    \label{geo_mag_scaling}
    g_M(x) \sim x^{-d_H \Delta}
\end{equation}
so that
\begin{equation}
    M(r;N) \sim r^{-d_H \Delta}.
\end{equation}
Figs.~\ref{geo_hist}, \ref{geo_mag}, and \ref{geo_spin_sum} show our finite
size scaling data 
for a representative value of $H=0.433$ and yield estimates for 
$1/d_H = 0.27 \pm 0.05 $ and $\Delta = 0.1 \pm 0.02$. 

The value for the fractal dimension, $d_H\approx 4$, is consistent
with its value on the sphere and compatible with the idea that the
fractal geometry of the disk far from the boundary is identical to the
quantum geometry of the sphere and the expectation that the influence
of the boundary decreases as one looks at spins further and further
into the disk.  This leads to the prediction that, in the continuum
limit, the bulk quantities should exhibit KPZ~\cite{KPZ:FSO}
behavior.

\begin{figure}\begin{center}
	\includegraphics[height=7cm]{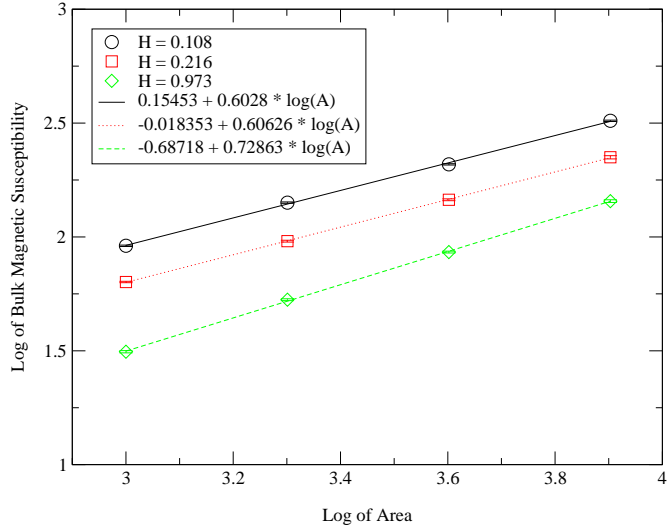}
	\caption{Scaling of bulk magnetic susceptibility with area for $H > 0$ at the tricritical point.}
	\label{tri_b_sus}
\end{center}\end{figure}

Specifically, we have examined the
scaling of the bulk magnetization and susceptibility at the
tricritical point.  Fig.~\ref{tri_b_sus} shows a plot of the logarithm
of the susceptibility versus the logarithm of the area for a variety
of boundary magnetic fields. The lines show good power law fits at a
variety of $H$ with exponents ranging from 0.6 to 0.7 which is in
approximate agreement with the KPZ prediction of $2/3$.

\begin{figure}\begin{center}
	\includegraphics[height=7cm]{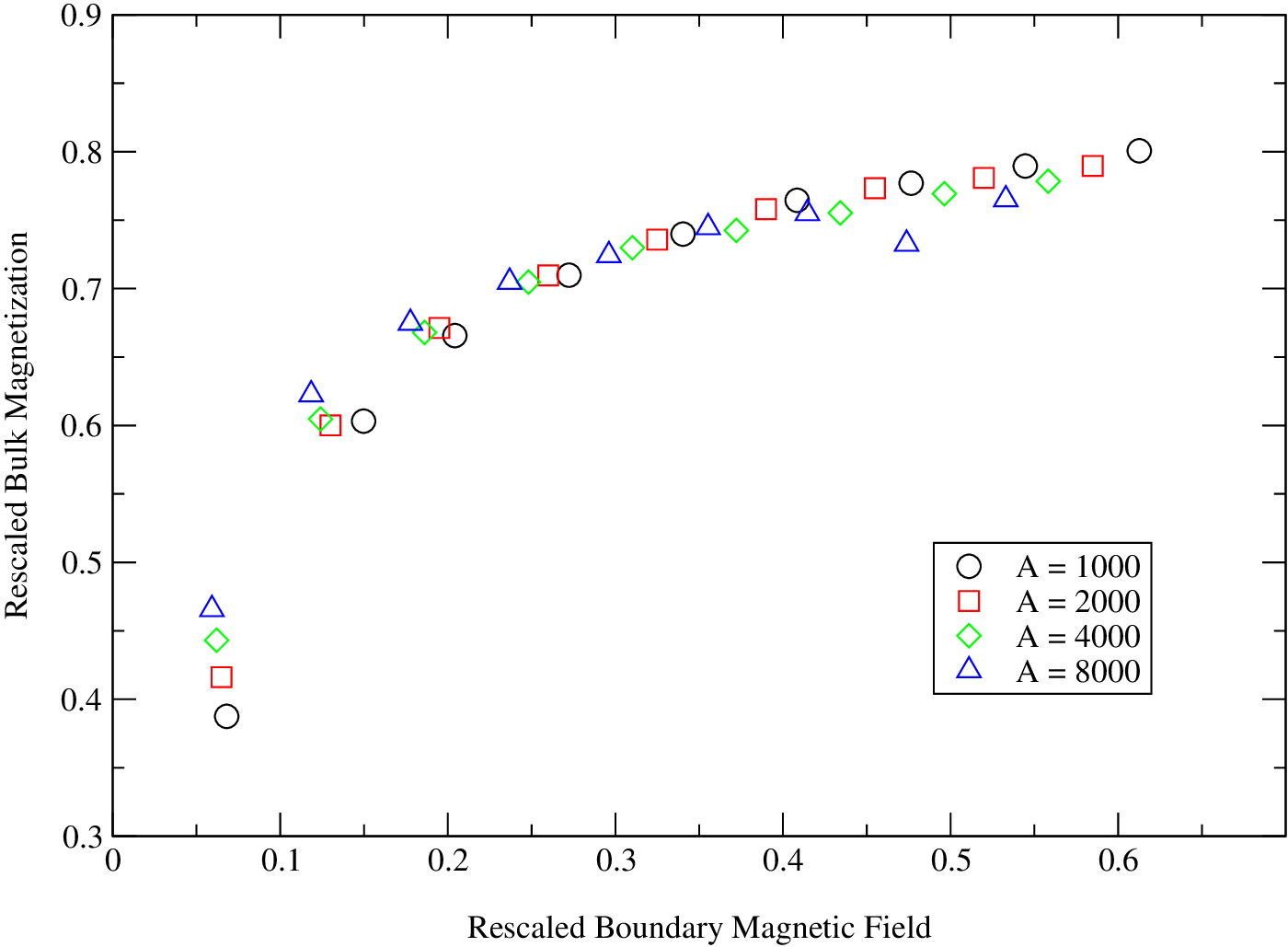}
	\caption{$\left<M_{\not\partial}\right> A^{0.1}$ vs. $H A^{-0.067}$ at the tricritical point.}
	\label{scaled_bulk_m}
\end{center}\end{figure}

The situation for the bulk magnetization is more complicated.  We
would expect that this would decrease with area according to the
simple power law $\left<M_{\not\partial}\right>\sim A^{-\Delta}$ with
$\Delta = \frac{1}{6}$ predicted by KPZ and consistent with
Eq.~\ref{M_bulk} using $l=\sqrt{A}$.
However, the finite size scaling result above as well
as fits of the same data as used in Fig.~\ref{bum_scaling} to powers
of area yield exponents closer to $0.1$.  This value
differs quite markedly from
the prediction.  This may reflect a finite size effect involving
the influence of spins close to the boundary where $d_H\approx 2$.
It is interesting to observe that the resulting exponent for $M(r;N)$,
$\Delta d_H \approx 0.37$ is near the value obtained in
\cite{COT:BFA,COT:TIM} of $0.33$ using $\Delta=\frac{1}{6}$ and $d_H =
2$.  From Eqs. \ref{M_bulk} and \ref{alpha},
\begin{equation}
    \left<M_{\not\partial}\right>A^{\frac{1}{6}} = {\left[\alpha(H)\right]}^{-\frac{1}{3}}.
\end{equation}
We hypothesize that the finite size effects can be taken into account
by generalizing this to
\begin{equation}
    \left<M_{\not\partial}\right> A^b = f(H A^a).
\end{equation}
(Note that rescaling $H$ with $A$ is equivalent to rescaling it with
$l$ since we have $l = \sqrt(A)$.)  Assuming $a < 0$ and expanding for
large $A$,
\begin{equation}
    \left<M_{\not\partial}\right> \approx f_0 A^{-b} + f_1 H A^{-b+a} + \frac{1}{2} f_2 H^2 A^{-b+2a} + ...
\end{equation}
Since we know that $\left<M_{\not\partial}\right>=0$ for zero $H$ we must have
$f_0=0$.  We found our data fit this well with $a =
-0.067 \pm 0.005$ and $b = 0.1 \pm 0.02$ (See Fig.~\ref{scaled_bulk_m}).
If our hypothesis is corect, then our data is consistent with
\begin{equation}
    \left<M_{\not\partial}\right> \sim A^{-0.167} \approx A^{-\frac{1}{6}}
\end{equation}
as predicted.

\begin{figure}\begin{center}
    \includegraphics[height=7cm]{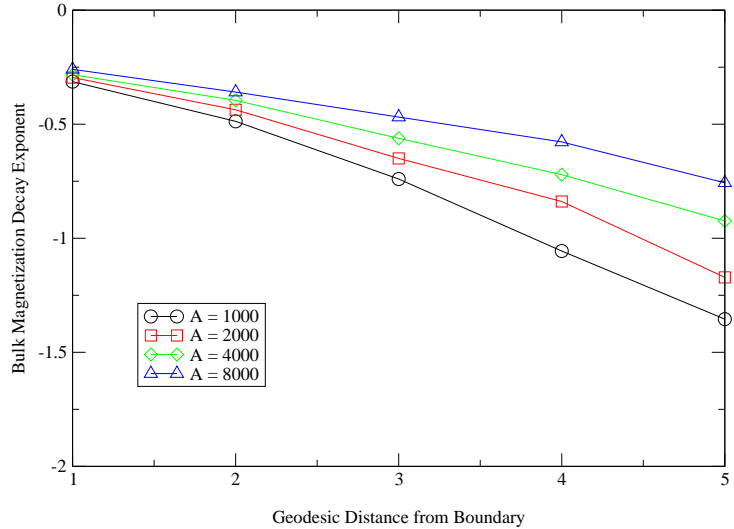}
    \caption{Exponents of bulk magnetization decaying with distance from the boundary at the tricritical point.}
    \label{first_five_m_decays}
\end{center}\end{figure}

To examine the scaling of the bulk magnetization further, we have calculated
\begin{equation}
    \frac{\log\left(\frac{M(r;N)}{M(r+1;N)}\right)}
             {\log\left(\frac{r}{r+1}\right)}
\end{equation}
which defines an effective (distance dependent) power law exponent for
this correlation function. Fig.~\ref{first_five_m_decays} shows
these exponents for $r = 1 \mbox{ to } 5$, averaged over several
values of $H$ from $0.24$ to $1.3$. 
The values for the different areas coincide for $d=1$ and then decrease 
at different rates as $r$ increases. The rate of decrease is smaller for larger
areas lending support to the hypothesis that a simple power law decay may
indeed describe the behavior in the thermodynamic limit.
The value of such a power law's 
exponent can be read off as approximately $-0.3$. This is consistent with
the finite size scaling results above. 

\section{Conclusion}

    We have mapped the phase diagram corresponding to a model of
Ising spins on a fluctuating geometry with disk topology. At zero
boundary magnetic field $H$ we find three phases separated by
critical lines on which either the magnetic or boundary length
susceptibilities are divergent with disk area. Only one of these
phase boundary lines corresponds to continuous transitions --- the
line separating ferromagnetic and paramagnetic phases on
geometries with small mean boundary length. The measured exponent
for this transition compares well with predictions from Liouville
theory. The third phase consists of elongated geometries whose
boundary length scales linearly with disk area. The continuum
limit is thought to reside at the junction of the these two
critical lines --- a tricritical point at which the spins 
display long range correlations and the mean boundary size
scales as the square root of the area $\left<l\right>\sim\sqrt{A}$.

We have tested the results of our simulations in two ways; by
comparison to hand counted amplitudes for small disks including the
Ising sector and by a direct computation of the critical exponents
governing the pure gravity disk partition function. The latter
constitutes the first numerical check on the values of these exponents
and was accomplished by a measurement of the baby universe
distribution (defined now for disk topology) and the density of 3-fold
coordinated vertices. In both cases good agreement with predictions
was obtained.

We have also computed the dependence of the bulk and boundary
magnetizations and susceptibilities on the boundary field $H$ at the
tricritical point. We find the boundary magnetization is independent
of boundary length or bulk disk area and varies smoothly with $H$. The
corresponding susceptibility approaches a constant as $H\to 0$ for
large area. Both of these conclusions are consistent with the
predictions of a matrix model calculation of a similar model by
Carroll, Ortiz and Taylor~\cite{COT:BFA,COT:TIM,COT:AGA} and at least
in the case of the susceptibility differ {\it qualitatively} from the
situation in flat space.

In the thermodynamic limit and at the tricritical point, a typical
bulk spin will be situated infinitely far from the boundary and
thus we might expect the critical behavior of the bulk
magnetization and susceptibility to be same as on a sphere {\it
independent} of the boundary field. Our numerical data for
the bulk susceptibility is indeed
consistent with this --- for a wide range of
fields $H$ we see a power law dependence on the area with
exponent in the range $0.6-0.7$ (the continuum/matrix model
prediction is $\frac{2}{3}$). The analysis of the magnetization data is
more complicated --- a naive power law fit of the area dependence at fixed
$H$ yields an exponent of approximately $0.1$ which appears quite
different from the predicted value of $\frac{1}{6}$. To
explain this we must
assume that the influence of the boundary spins induces large finite
size corrections in this exponent.

One conjecture made in the paper of Carroll et al. appears to be false
--- when we measure the bulk magnetization we find that it {\it
increases} with applied field --- contrary to the rather surprising
result of the matrix model calculation where a {\it decrease} of bulk
magnetization with increasing boundary field was found. In that case
it was conjectured that the physical interpretation of that result was
due to the appearance of a `neck' in the disk geometry which allowed
the average bulk spin to be pushed further out from the boundary with
increasing field. In our case we have verified that the opposite is
true --- the average bulk spin is drawn {\it closer} to the boundary
with increasing field. The explanation for this discrepancy
appears to stem from a lack of universality in the calculations --- our
lattice ensemble differs from the one used in the analytical
calculation and renders detailed comparison between the two
problematical. Specifically, the function $\alpha(H)$ which induces
this result is clearly not universal in character. Further tests of
universality obtained by placing the Ising spins on triangles are
currently underway.

On a more positive note we have checked another conjecture made in
the matrix model paper which speculates that the decay of the
boundary magnetization into its value in the bulk is governed by a
universal power of the geodesic distance to the boundary where the
power law exponent is related to the dressed magnetic exponent of 2-d
quantum gravity. We have measured both the number of
vertices and the magnetization as a function of geodesic distance from
the boundary and found that they exhibit finite size scaling forms for
geometries close to the tricritical point. Just as on the sphere it
appears that these functions reveal the existence of a single
length scale $\xi\sim A^{\frac{1}{d_H}}$ governing the behavior of
the average number of points and magnetization as we move from the
boundary into the interior of the disk. Indeed, the fractal
dimension $d_H$ appears to be close to its value on the sphere for
critical Ising spins coupled to gravity and lends support to the
idea that the dressing of Ising spin operators will just follow the
KPZ predictions. However, we find the extracted value of the bulk
magnetic operator differs from its exact value which we attribute to
large residual finite size corrections.

It would also be interesting to see which of
these qualitative features carried over to the case of the 
3-state Potts model coupled to 2-d gravity on the disk.

\section{Acknowledgements}
This research was supported by the Department of Energy, USA, under contract
number DE-FG02-85ER40237 and by research funds from Syracuse University.

\bibliographystyle{utphys}
\bibliography{bib}
\end{document}